\documentclass{article}

\usepackage[preprint]{cpal_2025}
\usepackage{wrapfig}

\renewcommand{\paragraph}[1]{\vspace{0.1em} \noindent \textbf{#1}~}

\usepackage{url}
\usepackage{hyperref}
\usepackage{booktabs}
\usepackage{csquotes}
\usepackage{nicefrac}
\usepackage{tikz}
\usetikzlibrary{arrows.meta,positioning,calc,shapes.geometric,patterns}
\usepackage{siunitx}

\DeclareSIUnit\flop{FLOP}
\title{LLMQ: Efficient Lower-Precision Pretraining \\ for Consumer GPUs}

\vspace{-2em}
\author{%
  Erik Schultheis\textsuperscript{1},
  ~Dan Alistarh\textsuperscript{1} \\
  \textsuperscript{1}IST Austria\\
  \texttt{first.last@ist.ac.at}
}

\begin{document}

\maketitle

\begin{abstract}
  We present LLMQ, an end-to-end CUDA/C++ implementation for medium-sized language-model
  training, e.g.\ 3B to 32B parameters, on affordable, commodity GPUs. These devices are characterized by low memory availability and
  slow communication compared to datacentre-grade GPUs. Consequently, we showcase a range of optimizations that target these bottlenecks, including activation checkpointing, offloading, and copy-engine based collectives. LLMQ is  able to train or fine-tune a 7B model on a single 16GB mid-range gaming card, or a 32B model on a workstation equipped with 4 RTX 4090s. Parallel training is achieved while executing a standard 8-bit training pipeline, without additional algorithmic approximations, and maintaining FLOP utilization of around 50\%. As such, the efficiency of LLMQ rivals that of production-scale systems on much more expensive cloud-grade GPUs. 
  We also present first results for training on the novel HP ZGX Spark device with a Blackwell-architecture GPU and unified memory. 
  Code is available at \url{https://github.com/IST-DASLab/llmq}.
\end{abstract}

\section{Introduction}
Large language models are extremely popular across many application areas, from code completion to personal assistants.
While training generalist models that excel in a wide variety of tasks requires compute resources out of reach for
anyone but large companies, many tasks can also be solved with smaller, highly specialized models. 
However, even such smaller models, e.g. in the 0.5 to 32B parameter range, are typically trained on large datacenter accelerators, which can be undesirable as it requires
uploading the training data to a third party, and can have high monetary cost. 

The alternative to this is \textit{local} training on \textit{user-owned hardware}, using more affordable gaming or even workstation GPUs. These accelerators can be 
significantly slower than their datacenter counterparts, but can be very competitive in terms of FLOPs per dollar.
From the systems perspective, these devices have two key drawbacks: 1) such GPUs have reduced  device memory, and 2) they also have much lower available bandwidth for communication (see Table~\ref{tab:gpu_comparison}).

\paragraph{Contributions} In this paper, we present the design and implementation of LLMQ, and end-to-end C++ framework for LLM pretraining and fine-tuning on commodity GPUs located in a single server node.
The basic version of LLMQ supports efficient bfloat16 training, and thus can be used across all recent GPU families, starting from the NVIDIA Ampere line (e.g. RTX 30xx). 
However, its key feature is in enabling highly-efficient and accurate FP8 training. For this, it uses dynamic tensor-level scaling, supported on both
Ada (RTX 40xx) and Blackwell (RTX 50xx) architectures. 
Our key technical contribution is the implementation of several strategies to alleviate memory bottlenecks:  

\begin{enumerate}
    \item To limit the cost of activations,
selective recomputation can be used, starting from only recomputing the non-matrix-multiplication layers, all the way to
recomputing full transformer blocks.
 \item Further peak memory reductions can be achieved by offloading the remaining residual to CPU RAM. Similarly, we show that optimizer states can be offloaded efficiently. 
 \item 
In a multi-GPU setup, LLMQ always shards optimizer states (ZeRO-1), and allows for sharding of model weights and gradients,
independently.
 \item As recent generation gaming GPUs are unable to communicate directly with each other over PCIe, we found that offloading
sharded parameters fully to the CPU does not increase the communication required during forward/backward passes while reducing
GPU memory usage further.
\end{enumerate}

By combining these optimizations, LLMQ achieves high throughput and utilization across consumer hardware configurations. On a workstation equipped with four RTX 4090s, the system attains a throughput of 7,800 tokens/second for a 14B parameter model, corresponding to 54\% Model FLOPs Utilization (MFU), leveraging LLMQ's custom communication backend. The framework successfully scales to a 32B parameter model on the same hardware, achieving 3,400 tokens/second at 51\% MFU , an efficiency rate significantly higher than that observed on professional L40S GPUs (29\% MFU). On a single RTX 4090, LLMQ can train models up to 14B parameters , while a 7B model runs at 4,300 tokens/second with 61\% MFU. Even on a constrained 16GB RTX 5060Ti, the system enables 7B model pretraining with a remarkable 70\% MFU. We also present results on the recent NVIDIA DGX Spark hardware.

\vspace{-1em}
\section{Background}
\vspace{-0.5em}
A number of system improvements have been devised to mitigate the large memory costs of LLM training. Four
main strategies are used: Sharding, recomputation, offloading and quantization.

Sharding, that is, distributing large tensors across multiple devices, is the key optimization in the well-established 
ZeRO~\citep{rajbhandari2020zero} optimization, in which optimizer states, and potentially also gradients and weights,
are spread out evenly across the devices. While sharding of optimizer states can be done for free, gradient sharding requires
additional communication if used together with gradient accumulation, and weight sharding always introduces additional
transfers.
Instead of having each worker process every layer of the network, it is also possible to distribute network layers across workers,
in a \emph{pipeline-parallel} setup~\citep{NEURIPS2019_093f65e0}.
Such a setup trivially avoids having to communicate model weights or gradients, but instead has to employ sophisticated 
scheduling to prevent idle workers due to pipeline bubbles~\citep{qi2024zero}.
For very long sequences, it might even be necessary to split the sequences across nodes in \emph{context parallism}~\citep{liu2024ringattention}.
For large scale trainings, these different forms of paralellism are combined into multidimensional parallelism~\citep{korthikanti2023reducing,ultrascale_playbook}.

To address the memory consumption of activation memory in particular, it is possible to recalculate a subset of activations during the backward pass
instead of keeping them in memory~\citep{chen2016training,korthikanti2023reducing}. A famous example of this technique is the avoidance of squared memory cost in attention 
layers~\citep{dao2022flashattention}.

As a last resort, one can also move large allocations down in the memory hierarchy into slower but more abundant storage, 
from GPU to CPU memory~\citep{ren2021zero} or even to NVME~\citep{rajbhandari2021zero}.

Finally, memory can also be saved by switching to lower-precision datatypes. For example, by compressing the moment
buffers of Adam~\citep{kingma2014adam} to use only 8 bits per element~\citep{dettmers2022bit}. By using mixed-precision
training~\citep{micikevicius2018mixed}, the activations can be compressed down to 16 bit or even 8 bit~\citep{micikevicius2022FP8formatsdeeplearning}
floating-point numbers. However, while the reduced dynamic range of 16 bit can be compensated for by either switching to
a BF16 representation with more exponent bits \citep{Kalamkar2019ASO}, or by introducing \emph{loss scaling}, the further
reduction to eight bits requires more sophisticated techniques. First, a combination of two different FP8 types, E5M2 and E4M3, 
with 5 and 4 exponent bits, respectively, can be used~\citep{sun2019hybrid}, depending on whether accuracy or range are more
important for a specific tensor. Second, additional scale factors can be introduced that scale values into the representable 
range before quantization. Such scale factors can be applied at the level of an entire tensor, either using its current
values (just-in-time scaling) or based on previous steps' values (delayed scaling). More fine-grained scaling is also
possible, giving one scale per token or channel, or even small sub-blocks of the tensor~\citep{rouhani2023microscaling}.
Scaling could either be handled such to ensure that no overflows occur (abs-max scaling), or accept a small amount of clipping
for better fidelity~\citep{panferov2025quest}.

In addition to memory savings, for low-precision formats that are natively supported in hardware, there are also
 computational advantages. Smaller data formats mean less data movement, and fewer transistors and energy expenditure required 
to handle fused multiply-add operations~\citep{patterson2016computer}. This gets amplified by specialized hardware units, such as tensor cores~\citep{choquette2020nvidia},
which operate on these formats.
To benefit, any scaling that gets applied along the inner dimension of the matrix multiplication either needs native hardware support, only available
on recent Blackwell GPUs~\citep{blackwell}, or requires software emulation with overheads~\citep{deepseekai2025deepseekv3technicalreport}.

\section{Implementation}
In this section, we describe the implementation of LLMQ, starting from overall design decisions, to the
optimizations necessary for efficiently running larger models on a single and on multiple GPUs.

\paragraph{Overview} LLMQ is implemented in C++ and CUDA. For matrix multiplications and attention, we use the available implementations in
cuBLAS and cuDNN, respectively; all other kernels are implemented by us, as part of the project, and are provided as open-source.
All memory allocations in LLMQ happen at program startup. This means that if the program does not run out of memory
before the first step, it will never run out of memory. (In extreme circumstances (<50MiB free), it is possible to get OOM when the GPU runs out of memory to load the kernels during the first step.)

We provide both a pure BF16 implementation, and mixed BF16-FP8 training. In order to support older GPU cards, we use just-in-time
tensor-level absmax-scaling for conversion from BF16 to FP8, that is, we first determine the largest absolute value
in each tensor, and then rescale so that this is mapped to the largest representable value. Compared to delayed scaling,
this requires an additional kernel call (due to the global reduction), but it guarantees that no value will be clipped
even if tensor statistics change rapidly.
Main transformer matmuls in forward and backward are calculated in FP8, whereas nonlinearities, SDPA, the embeddings and LM-head,
as well as gradient accumulation remain in BF16. The latter ensures that many steps of gradient accumulation can be performed
without catastrophic cancellation in the summation.

\paragraph{Reproducibility} To achieve reproducible (on the same hardware, with the same software) training runs, LLMQ only uses bitwise-deterministic
kernels. This is not a problem during the forward pass, but it means that for reductions required in the backward pass,
instead of atomic additions, an intermediate buffer to store local result, and a second kernel that handles the global
reduction, are needed.
Particularly challenging in that regard is the backward pass for the embedding layer: If tokens get distributed \textquote{randomly}
to different thread blocks, then an intermediate buffer of size \texttt{embeddings}$\times$\texttt{num-blocks} would be needed.
Instead, as implemented in \texttt{llm.c} (\url{https://github.com/karpathy/llm.c}), the token indices are first sorted and partitioned on the CPU,
so that each thread block can focus on a small subset of tokens. The sorting process can overlap with the regular backward pass,
and thus does not introduce additional latency.
In cases where random decisions need to be taken inside GPU kernels (e.g., for stochastic rounding), we use counter-based
generators to draw deterministic pseudo-random numbers without requiring an internal state.

\paragraph{Multi-GPU Support} Within a single node, there are two main approaches for handling multi-gpu support.
One is to use one process per GPU, the other to use multiple threads in a single 
process. Multiprocessing can be beneficial in avoiding the global interpreter lock in python,
which is of no concern to use, and scales beyond a sinlge node. On the other hand,
in a multi-threading setup, one can exploit the shared address space which allows
direct GPU-to-GPU memcpy without having to resort to IPC handles. While we do support
both options, the focus is on the multi-threaded setup and its direct communication options.

When running on multiple GPUs, LLMQ \emph{always} shards optimizer states (aka ZeRO-1).
This is strictly better than traditional DDP with replicated optimizer states, as this
leads to reduced memory consumption without increasing the amount of communication.

To avoid unnecessary round trips to device memory, we fuse all successive operations
that are not either a global reduction or involve a matrix multiplication.
In particular, this means that all our non-linearity operators have an additional
output parameter that returns the abs-max of its result. Further, RMS-norm and
residual-stream addition are handled in a joint kernel, which then also returns
the abs-max of the RMS-norm.
As in FP8 on consumer cards, gemm only supports the \texttt{TN} (transpose:non-transpose) layout,
we need to handle the transposes manually, including a fused transpose+quantize kernel.
Finally, we fuse the forward and backward pass of the cross-entropy loss into a single kernel~\citep{renee_2023,hsu2025ligerkernel}, avoiding the need to materialize
a huge per-token loss tensor.

\subsection{Training on a Single Mid-level commodity GPU: RTX 5060Ti}

Next, we will consider the optimizations necessary to enable training a 7B model
with only 16GB of device memory. On a single 16GB GPU, the implementation described above allows training $0.5B$ parameter models at batch size 6,
but runs out of memory for $1.5B$.

\paragraph{Activation checkpointing}
To alleviate the memory pressure coming from the activations stored for the backward
pass, it is possible to only keep a select subset of activations and recompute the
others during backward.
If only a moderate amount of memory is needed, it might be sufficient to only
recompute the non-gemm values, that is, swiglu and rmsnorm, but for drastic reductions
the entire transformer block needs to be recomputed, only keeping the residual of
the feed-forward part. As memory capacities vary significnatly between commodity cards,
it is not possible to decide the optimal trade-off a-priori; therefore, LLMQ allows for
selective recomputation that covers the full range.

In addition to preserving the feed-forward residual, we also always keep small statistics
tensors from the forward pass. That means that during recomputation, we do not have to
determine the absmax again. In particular, with the absmax known, there is no longer a need
for a global reduction, and the quantization can be fused into the nonlinearity.

\paragraph{Reduced-precision optimizer states}
A second large contributor to memory consumption are the optimizer states. By default,
these are kept in fp32 precision, which for AdamW corresponds to 8 bytes per parameter,
or 12GB in total for a 1.5B model.
This shrinks by a factor of two when switching to BF16 to represent momentum and variance.
To ensure unbiased values, conversion of fp32 to bf16 is handled by stochastic rounding.
Note that we keep master copies of parameters only in bf16, too.

\paragraph{Offloading}
As the optimizer states still make up a large fraction of total device memory, further memory
cost reduction can be achieved by offloading them to host memory. Here we have two options:
Either just keep them in page-locked (pinned) memory and rely on the GPUs zero-copy ability
to directly read from host, or allocate smaller buffers on the GPU to do explicit double-buffering.
Interestingly, we found that zero-copy gave bad performance (i.e., low utilization of PCIe bandwidth)
on gaming GPUs (5060Ti, 4090 but worked well on the more high-end cards (L40s), whereas the situation
was reversed for double-buffering.
Consequently, we recommend testing both options an the system in question and picking the faster one.

With these improvements, we can run 1.5B with batch-size 2 (no offloading) or 12 (offloading both $m$ and $v$).
To enable the 3B model, we additionally need to offload the 16-bit master copies of the matrix parameters, in which
case a batch size of 8 can be achieved.

Finally, we can also offload the last remaining residuals to host memory, which increases the maximum batch size to 10.
At this point, activation memory is dominated by the cost of a single layer.

\paragraph{Chunking}
In particular, it turns out that two tensors grow very large as the batch size increases. The first, unsurprisingly. are the logits,
due to the large vocabulary dimension. Somewhat unexpectedly, the second is the workspace needed for cuDNN to enable deterministic 
backward for flash-attention.

Both of these can be solved by chunking the calculation. For the logits, this means that we split the pre-lmhead embeddings
into small chunks, and do a full forward-backward over each chunk, writing the derivative of the embeddings into the corresponding slice
and accumulating the gradients of the lm-head weights\citep{hsu2025ligerkernel}.
For attention, it is even simpler, because there are no trainable parameters: Simply iterate over slices of input and output, calling the cuDNN
kernel multiple times with a smaller workspace.

By fixing these bottlenecks, we can increase the batch size for the 3B model all the way to 24.

\paragraph{Enabling 7B models on 16GB VRAM}
The preceding optimizations allow increasing the micro-batch size significantly. This, in turn, means that the amount of computation done in each forward/backward pass increases, and consequently, more communication can be hidden without affecting end-to-end latency.

This gives us the opportunity to offload even the gradient buffers to host memory.\footnote{As explained in the next section, we only offload transformer blocks, not lm-head and embeddings}
Thus, the device memory consumption reduces to that of two layers; one for active computations, and one for double-buffering data transfers. As such, even a 7B model can be run with micro-batch size 16.
As this point, however, even a high-end gaming PC will reach its limits of available host memory: A 7B model needs approximately $3\times14$GB for the optimizer state, another $7$GB for the quantized weights, and $5$GB for offloaded residuals, a total of 54GB of memory. While it might be possible to offload one more level, from host memory to an NVME~\citep{rajbhandari2021zero}, doing so will change the speed of transfers by another order of magnitude, no longer hidden behind the computations.

To achieve further total memory reductions, one would need to either switch to even lower bit-width optimizers~\citep{dettmers2022bit,xi2025coat,huang2025stable}, optimizers with smaller buffers~\citep{modoranu2024microadam,zhao2024galore,shazeer2018adafactor}, or parameter-efficient fine-tuning~\citep{hu2022lora}. While possible and interesting, these optimizations perform algorithmic approximations, and are therefore orthogonal to our work. 

\subsection{Multi-GPU Training on a 4x RTX 4090 Workstation}
In this section, we consider a workstation with 4$\times$RTX 4090 GPUs. This allows for sharding of optimizer states, weights, and gradients, before we need to resort to CPU offloading.

\paragraph{Weight caching on host}
However, on recent consumer devices, the GPUs cannot communicate directly to each other via PCIe, and instead need to go through the host. As such, placing weights in host memory actually reduces the required communication: They need to be sent from GPU to host during the first forward pass after each optimizer step, but the backward pass, and all subsequent forward passes for gradient accumulation, can use the values cached on the host.

This also means that, contrary to traditional ZeRO~\citep{rajbhandari2020zero} levels, one should enable sharded model weights \emph{before} enabling sharded gradients. This is especially true in FP8 training, when model weights are gathered in FP8 but gradients are communicated in BF16.

\paragraph{Imbalances of LM-head and embeddings}
In particular, due to the large vocabulary dimension, both compute and communication cost for the LM-head exceed that of a regular transformer block. For this reason, we forgo sharding the LM-head/embedding, instead replicating them across each worker. Thus, their gradients need only be synchronized at the last gradient accumulation step. By having the LM-head in a separate buffer than the double-buffered transformer block, we further ensure that during that last backward pass,
sending its gradient can be overlapped with computing the gradients for the last two transformer blocks, whose weights are still available locally from the preceding forward pass.
Additional overlap can be achieved by scheduling the two backward matrices of the LM-head gradient such that the one that calculates the weight gradient is handled first, and the communication can be overlapped with the input gradient calculation. However, the effectiveness of this optimization diminishes with increased chunking, as sending can only commence in the last chunk.

Unfortunately, for the token embeddings, the next required operation is the global norm reduction of the gradients, so there is no way to hide that communication latency.

\paragraph{cudaMemcpy-based communication}
When running \texttt{all-gather} (fetching model weights) and \texttt{reduce-scatter} (accumulating gradients) colletives with \texttt{nccl}, we observed that the PCIe link utilization was quite low.
On the other hand, memory transfers handled by the GPUs copy-engines, as initiated by \texttt{cudaMemcpy}, do achieve much better utilization, and do not require reserving any SMs for communication kernels.

\begin{figure}
    \centering
    \begin{tikzpicture}[
    chunk/.style={rectangle, draw, thick, minimum width=0.75cm, minimum height=0.4cm, font=\tiny},
    accum/.style={rectangle, draw, thick, minimum width=0.75cm, minimum height=0.4cm, font=\tiny, pattern={north east lines}},
    arrow/.style={-Stealth, thick},
    label/.style={font=\footnotesize\sffamily},
    phase/.style={font=\small\bfseries\sffamily}
]

\node[label] at (0, -0.0) {$W_0$:};
\node[label] at (0, -1.5) {$W_1$:};
\node[label] at (0, -3) {$W_2$:};

\node[phase] at (1.5, 0.5) {Local Aggregation};

\node[chunk, fill=blue!30] (p1w0c0) at (1, 0.0) {$G_0^{0}$};
\node[chunk, fill=red!30] (p1w0c1) at (1.85, 0.0) {$G_1^{0}$};
\node[chunk, fill=green!30] (p1w0c2) at (2.7, 0.0) {$G_2^{0}$};
\node[accum, pattern color=blue!50] (p1w0acc) at (1, -0.7) {$G_0^{0}$};

\node[chunk, fill=blue!50] (p1w1c0) at (1, -1.5) {$G_0^{1}$};
\node[chunk, fill=red!50] (p1w1c1) at (1.85, -1.5) {$G_1^{1}$};
\node[chunk, fill=green!50] (p1w1c2) at (2.7, -1.5) {$G_2^{1}$};
\node[accum, pattern color=red!50] (p1w1acc) at (1.85, -2.2) {$G_1^{1}$};

\node[chunk, fill=blue!70] (p1w2c0) at (1, -3) {$G_0^{2}$};
\node[chunk, fill=red!70] (p1w2c1) at (1.85, -3) {$G_1^{2}$};
\node[chunk, fill=green!70] (p1w2c2) at (2.7, -3) {$G_2^{2}$};
\node[accum, pattern color=green!50] (p1w2acc) at (2.7, -3.7) {$G_2^{2}$};

\draw[arrow, blue!70, dashed] (p1w0c0.west) to[bend right=40] (p1w0acc.west);
\draw[arrow, red!70, dashed] (p1w1c1) -- (p1w1acc);
\draw[arrow, green!70, dashed] (p1w2c2.east) to[bend left=40] (p1w2acc.east);

\node[phase] at (5, 0.5) {Round 1};

\node[chunk, fill=gray!50] (p2w0c0) at (4.5, 0.0) {$G_0^{0}$};
\node[chunk, fill=red!30] (p2w0c1) at (5.35, 0.0) {$G_1^{0}$};
\node[chunk, fill=green!30] (p2w0c2) at (6.2, 0.0) {$G_2^{0}$};
\node[accum, pattern color=blue!50] (p2w0acc) at (4.5, -0.7) {$G_0^{0}$};

\node[chunk, fill=blue!50] (p2w1c0) at (4.5, -1.5) {$G_0^{1}$};
\node[chunk, fill=gray!50] (p2w1c1) at (5.35, -1.5) {$G_1^{0}$};
\node[chunk, fill=green!50] (p2w1c2) at (6.2, -1.5) {$G_2^{1}$};
\node[accum, pattern color=red!50] (p2w1acc) at (5.35, -2.2) {$G_1^{1}$};

\node[chunk, fill=blue!70] (p2w2c0) at (4.5, -3) {$G_0^{2}$};
\node[chunk, fill=red!70] (p2w2c1) at (5.35, -3) {$G_1^{2}$};
\node[chunk, fill=gray!50] (p2w2c2) at (6.2, -3) {$G_2^{1}$};
\node[accum, pattern color=green!50] (p2w2acc) at (6.2, -3.7) {$G_2^{2}$};

\draw[arrow, blue] (p2w2c0.west) .. controls (3.5,-3) and (3.5,-3) .. (3.5, -1.5) .. controls (3.5,-0.5) and (3.5,-0.0) .. (p2w0c0.west);
\draw[arrow, red] (p2w0c1.south) -- (p2w1c1.north);
\draw[arrow, green] (p2w1c2.south) -- (p2w2c2.north);

\node[phase] at (8.5, 0.5) {Round 2};

\node[chunk, fill=blue!70] (p3w0c0) at (8, 0.0) {$G_0^{2}$};
\node[chunk, fill=gray!50] (p3w0c1) at (8.85, 0.0) {$G_0^{1}$};
\node[chunk, fill=green!30] (p3w0c2) at (9.7, 0.0) {$G_2^{0}$};
\node[accum, pattern color=blue!50] (p3w0acc) at (8, -0.7) {$G_0^{0}$};

\node[chunk, fill=blue!50] (p3w1c0) at (8, -1.5) {$G_0^{1}$};
\node[chunk, fill=red!30] (p3w1c1) at (8.85, -1.5) {$G_1^{0}$};
\node[chunk, fill=gray!50] (p3w1c2) at (9.7, -1.5) {$G_1^{2}$};
\node[accum, pattern color=red!50] (p3w1acc) at (8.85, -2.2) {$G_1^{1}$};

\node[chunk, fill=gray!50] (p3w2c0) at (8, -3.0) {$G_2^{0}$};
\node[chunk, fill=red!70]   (p3w2c1) at (8.85, -3.0) {$G_1^{2}$};
\node[chunk, fill=green!50] (p3w2c2) at (9.7, -3.0) {$G_2^{1}$};
\node[accum, pattern color=green!50] (p3w2acc) at (9.7, -3.7) {$G_2^{2}$};

\draw[arrow, green] (p3w0c2.south) .. controls (9.7, -1.12) .. (9.0, -1.12) .. controls (7.2,-1.12) .. (7.2, -2) .. controls (7.2,-3.0) .. (p3w2c0.west);
\draw[arrow, blue] (p3w1c0.north)  .. controls (8, -1.02) .. (8.5, -1.02) .. controls (8.85,-1.02) .. (p3w0c1.south);
\draw[arrow, red] (p3w2c1.north) .. controls (8.85, -2.6).. (9.0, -2.6) .. controls (9.7,-2.6) .. (p3w1c2.south);

\node[phase] at (12, 0.5) {Reduction};

\node[chunk, fill=blue!70] (p4w0c0) at (11.5, 0.0) {$G_0^{2}$};
\node[chunk, fill=blue!50] (p4w0c1) at (12.35, 0.0) {$G_0^{1}$};
\node[chunk, fill=green!30] (p4w0c2) at (13.2, 0.0) {$G_2^{0}$};
\node[accum, pattern color=blue!50] (p4w0acc) at (11.5, -0.7) {$\hskip-2pt\smash{\sum_{\!j}}\hskip-1ex G_0^{j}\hskip-2pt$};

\node[chunk, fill=blue!50] (p4w1c0) at (11.5, -1.5) {$G_0^{1}$};
\node[chunk, fill=red!30] (p4w1c1) at (12.35, -1.5) {$G_1^{0}$};
\node[chunk, fill=red!70] (p4w1c2) at (13.2, -1.5) {$G_1^{2}$};
\node[accum, pattern color=red!50] (p4w1acc) at (12.35, -2.2) {$\hskip-2pt\smash{\sum_{\!j}}\hskip-1ex G_1^{j}\hskip-2pt$};

\node[chunk, fill=green!30] (p4w2c0) at (11.5, -3.0) {$G_2^{0}$};
\node[chunk, fill=red!70] (p4w2c1) at (12.35, -3.0) {$G_1^{2}$};
\node[chunk, fill=green!50] (p4w2c2) at (13.2, -3.0) {$G_2^{1}$};
\node[accum, pattern color=green!50] (p4w2acc) at (13.2, -3.7) {$\hskip-2pt\smash{\sum_{\!j}}\hskip-1ex G_2^{j}\hskip-2pt$};

\draw[arrow, blue, dashed] (p4w0c0.west)  to[bend right=35]  (p4w0acc.west);
\draw[arrow, blue, dashed] (p4w0c1.south) to[bend left=35] (p4w0acc.east);

\draw[arrow, red, dashed] (p4w1c1)        -- (p4w1acc);
\draw[arrow, red, dashed] (p4w1c2.south)  to[bend left=35] (p4w1acc.east);

\draw[arrow, green, dashed] (p4w2c0.south) to[bend right=20] (p4w2acc.west);
\draw[arrow, green, dashed] (p4w2c2.east)  to[bend left=35] (p4w2acc.east);

\draw[-Stealth, thick] (0, -4.2) -- (3, -4.2);
\draw[-Stealth, thick] (3, -4.5) -- (10.5, -4.5);
\draw[-Stealth, thick] (10.5, -4.2) -- (14, -4.2) node[right, label] {};
\node[label, below] at (1.5, -4.2) {SM; main stream};
\node[label, below, align=center] at (6.625, -4.5) {CE; copy stream.};
\node[label, below] at (12.25, -4.2) {SM; main stream};

\end{tikzpicture}
    \caption{Memcpy-based reduce-scatter. After the local shard has $G_i^i$ of worker $W_i$ has been added to the local accumulator (striped), in each round of communication each worker has exactly one chunk that is not needed anymore, that can be used as a destination buffer for memcpy. At the end, each worker has the corresponding chunks of the other workers, and can accumulate them. The communication phase can be purely handled by the GPU's copy engine (CE), leaving the streaming multiprocessors (SM) available to run the backward pass of the next transformer block.}
    \label{fig:memcpy-reduce}
\end{figure}
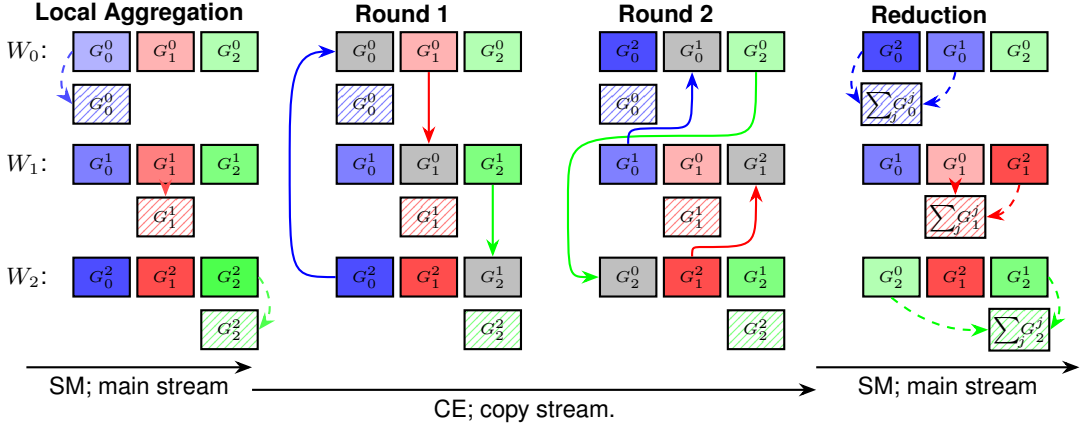

For \texttt{all-gather}, replacing \texttt{nccl} kernels with copies is trivial, as gathering only moves bytes around.
However, \texttt{reduce-scatter} mixes arithmetic and data movement, so it cannot be implemented by copy-engines alone. Therefore, we split the implementation into three phases. First, the local chunk for the current layer is aggregated to the sharded gradient buffer. Then, as this chunk of memory is no longer needed, it can be used as a scratch buffer for receiving chunks from the other workers. This happens in round-robin fashion, that is, after having received chunk and sent one chunk, another part of the memory is unused and can be used as buffer.
The copying operations do not need any multiprocessors, and thus can be run perfectly parallel to the backward of the next transformer layer. Only after that transformer layer has finished do we need to synchronize and wait for the transfers to be done. At that stage, the preceding layers gradient buffer contains the gradients for the local shard that have been gathered from the other workers, and we can run the reduction operations, adding them with stochastic rounding. This  is illustrated in \autoref{fig:memcpy-reduce}.

\paragraph{Multi-threaded multi-GPU and deadlocks}
While the multi-threading-based parallelization makes implementing memcpy communication much easier, we found it to have some drawbacks when involving nccl collectives. Specifically, the collective operations involve a global barrier that all GPUs must reach before any can make progress. 
We observed that, in some cases, this could lead to deadlocks. While we do not know the exact cause, we hypothesize the following: There is some per-process resource that gets filled up as GPU operations are enqueued. If one GPU is faster than another, it could enqueue the collective, and then continue enqueueing further kernels until that resource is exhausted. Because of the barrier in the collective, the GPU cannot execute any more kernels to make space for issuing new operations, and the other GPU cannot issue kernels it needs to reach the collective because there is no space.
By placing CPU-side synchronization (that is, the CPU threads are synchronizing among each other, but \emph{not} with the GPU), we can prevent new kernels getting submitted until every worker has issued the collective, which fixes the deadlocks.

\section{Benchmarks}

In this section, we provide an overview of the speed achievable in different scenarios.
\autoref{tab:speed-benchmark-single} shows the tokens per second and model-flops-utilization for differently-sized models, trained at a per-step batch size of 500k, on a single GPU. For each configuration, the combination of offloading/recomputation/micro-batch size that leads to the highest throughput was chosen.
Model Flops Utilization (MFU) is calculated taking into account the mixed precision nature of the implementation, that is, we calculate the amount of floating-point operations to be done in each precision, divide by the device's peak rate, and get a lower bound for the achievable duration. The ratio of achievable duration to actual timing is presented in the MFU columns.

\begin{table}[]
    \centering
    \caption{Training speed and utilization on a single gpu. Sp denotes speed-up of FP8 over BF16.}
    \small 
    \begin{tabular}{r|rr|rr|r|rr|rr|r|r}
        \toprule
             & \multicolumn{5}{c|}{RTX 5060Ti} & \multicolumn{6}{c}{RTX 4090} \\[0.2ex]
        Size & \multicolumn{2}{c|}{FP8} & \multicolumn{2}{c|}{BF16} & Sp &
        \multicolumn{2}{c|}{FP8} & \multicolumn{2}{c|}{BF16} & Sp & LF\\
             & TPS   & MFU     & TPS & MFU && TPS   & MFU     & TPS & MFU && TPS
         \\ \midrule
        0.5B & 16.5k & 70\%    & 13.0k    & 85\%    & 27\%    & 47k   & 58\%    & 39k   & 73\%    & 20\% & 30.4k \\  
        1.5B & 5.7k  & 67\%    & 3.9k     & 78\%    & 46\%    & 20k   & 69\%    & 14k   & 81\%    & 42\% & 9.5k \\
        3B   & 3.1k  & 69\%    & 2.0k     & 80\%    & 55\%    & 10.6k & 68\%    & 7.0k  & 80\%    & 51\% & 5.4k \\
        7B   & 1.4k  & 70\%    & 0.9k     & 79\%    & 55\%    & 4.3k  & 61\%    & 2.7k  & 71\%    & 59\% & 2.5k \\ 
        14B  & ---   & ---     & ---      & ---     & ---     & 2.0k  & 55\%    & 1.3k  & 59\%    & 54\% & 1.2k \\ 
        \bottomrule
    \end{tabular}
    \label{tab:speed-benchmark-single}
\end{table}

\begin{table}
    \centering
    \caption{Training speed and utilization on multiple GPUs. Sp denotes speed-up of FP8 over BF16.}
    \small 
    \begin{tabular}{r|rr|rr|r|rr|rr|r|r}
        \toprule
             &\multicolumn{5}{c|}{$4\times$L40S} & \multicolumn{6}{c}{$4\times$RTX 4090} \\[0.2ex]
        Size & \multicolumn{2}{c|}{FP8} & \multicolumn{2}{c|}{BF16} & Sp &
        \multicolumn{2}{c|}{FP8} & \multicolumn{2}{c|}{BF16} & Sp & LF \\
             & TPS   & MFU     & TPS & MFU && TPS   & MFU     & TPS & MFU & TPS
         \\ \midrule
        0.5B & 191k  &  27\%   &  185k   &  37\%   &  3\% & 181k  & 55\% & 154k  & 71\% & 18\% & 123k \\  
        1.5B & 81k   &  31\%   &  68k    &  44\%   & 30\% &  71k  & 60\% & 52k   & 75\% & 36\% & 41k \\
        3B   & 45k   &  32\%   &  36k    &  37\%   & 25\% &  38k  & 61\% & 26k   & 74\% & 46\% & 19k \\
        7B   & 21k   &  34\%   &  16k    &  38\%   & 31\% & 16.5k & 58\% & 10.9k & 69\% & 51\% & 6.9k \\ 
        14B  & 10k   &  31\%   &  7.1k   &  42\%   & 41\% & 7.8k  & 54\% & 5.2k  & 67\% & 50\% & 2.6k \\
        32B  & 4.2k  &  29\%   &  3.0k   &  40\%   & 30\% & 3.4k  & 51\% & 2.2k  & 65\% & 54\% & OOM \\   
        \bottomrule
    \end{tabular}
    \label{tab:speed-benchmark}
\end{table}

\begin{table}[ht]
    \centering
    \begin{minipage}{0.48\textwidth}
        \centering
        \caption{Training speed and utilization on a DGX Spark. Sp denotes speed-up of FP8 over BF16.}
        \begin{tabular}{r|rr|rr|r}
            \toprule
                 &\multicolumn{5}{c}{DGX Spark} \\[0.2ex]
            Size & \multicolumn{2}{c|}{FP8} & \multicolumn{2}{c}{BF16} & Sp\\
                 & TPS   & MFU     & TPS & MFU &
             \\ \midrule
            0.5B & 17k   &  28\%   &  17k  &  44\%   &  0\% \\  
            1.5B & 7.0k  &  33\%   & 6.5k  &  53\%   &  8\% \\
            3B   & 4.4k  &  39\%   & 3.6k  &  57\%   & 22\% \\
            7B   & 2.4k  &  47\%   & 1.7k  &  63\%   & 41\% \\ 
            \bottomrule
        \end{tabular}
        \label{tab:speed-benchmark-spark}
    \end{minipage}
    \hfill 
    \begin{minipage}{0.48\textwidth}
        \centering
        \caption{Comparison between datacentre and gaming GPUs}
        \small 
        \begin{tabular}{l|ccc}
            \toprule
                                     & H100   & RTX 4090 & Ratio \\ \midrule
            BF16 [TFLOP/s]           & 989.4  & 165.2    & $6   \times$ \\
            Memory [GB]              & 80     & 24       & $3.3 \times$ \\ 
            Bandwidth [TB/s]         & 3.3    & 1        & $3.3 \times$ \\
            Cost [\$]                & 30k    & 2k       & $15 \times$  \\
            Power [W]                & 700    & 450      & $1.5 \times$ \\
            Communication            & NVLink & PCIe 4.0 & --           \\
            Bandwidth [GB/s]         & 900    & 64       & $14$         \\ \bottomrule
        \end{tabular}
        \label{tab:gpu_comparison}
    \end{minipage}
\end{table}

\paragraph{Impact of FP8}
The relative speed-up due to lower precision is shown in the SP column, and reaches about $50\%$ for sufficiently large models. For very small models, non-gemm operations contribute significantly to the total running time, so speeding up the GEMMs only provides limited improvements. Yet, FP8 comes with additional overheads due to dynamic quantization, and the transposes required in the backward pass.
As the model size increases, compute is dominated by matrix math, and FP8 can lead to significant boosts.
Due to the fixed batch size, for very large models, the optimizer step starts taking a noticeable fraction of the time.

Notice that part of the computations are still done in 16-bit precision.
For the 7B model, the operations break down to \num{39.2e9} ops in FP8 for the linear operations inside the transformer blocks, \num{3.3e9} operations in bf16 for the LM-head, and \num{0.6e9} operations in BF16 for attention. Consequently, even if there were no quantization-related overheads, we could expect FP8 to only bring a speed-up of 90\%. Despite falling short of the theoretical maximum, the achieved speed-ups are generally in line with existing FP8 implementations \citep{fp8lm,hernandez-cano2025towards}.

Interestingly, significant benefits from FP8 generally do \emph{not} come from enabling a larger batch size due to lower activation memory requirements. For small models, overheads limit the effectiveness of FP8, and for large models, activation recomputation means that we do not store the activations that would be compressed to FP8 (e.g., residuals remain in BF16); even worse, FP8 requires additional buffers for transposes and quantization, thus actually using \emph{more} memory when entire transformer blocks are recomputed.

Comparing the RTX 5060Ti and the 4090, we generally get slightly better utilization on the smaller card. While this is expected for small models, at first it might be
surprising for larger models; however, the 5060Ti has only about $\nicefrac{1}{3}$ of the FLOP/s of the 4090, yet provides the same PCIe bandwidth, so even though it
requires more offloading, it is also better able to hide the resulting latencies.

\paragraph{Llama-Factory Comparison} Finally, the \textquote{LF} column show the speed of LLama-Factory (LF)\citep{zheng2024llamafactory} on the 4090. For small models, the llmq implementation is significantly faster, but for large models, the difference almost disappears. Interestingly, the strategies to achieve optimal performance at increasing model size differ between the two frameworks. For llmq, parts of the model are offloaded until at least a moderate batch size can be sustained; for LF, however, we found that, as soon as offloading is required, it is more efficient to do full offloading in order to support a very large batch size, than to do partial offloading at medium batch sizes.
We attribute this to the fact that the overheads associated with each forward/backward propagation are much lower in llmq than in LLama-factory.

Scaling up to multiple GPUs, \autoref{tab:speed-benchmark} shows training speeds on a system with 4 consumer GPUs (4090), compared against 4 professional-grade GPUs (L40s). We can see that, despite their nominally much faster peak FLOP/s, for small models the L40S are only a little faster than the 4090s, especially in FP8,
and even for large models, the gap is lower than expected, as the relative utilization on the L40S is considerably lower. This appears to be because the L40S run into
thermal and/or power throttling, never achieving actual peak performance, see also appendix~\ref{app:peak-flops}.

The comparison against LLama-factory shows the tremendous benefit conferred by efficient \texttt{memcpy}-based communication. For small models that do not
require parameter sharding, there is only a minor performance gap (52k vs 41k for 1.5B parameters in BF16), but at the largest scale supported by LF, 14B, the llmq implementation is twice as fast.

\begin{table}
    \centering
    \caption{Effect of \texttt{nccl} collectives compared to \texttt{cudaMemcpy}-based communication of consumer (RTX 4090) and professional (L40S) cards for the 14B model.
    Comparing \texttt{nccl} (None) with enabling \texttt{memcpy} only for one of the collectives (Gather, Scatter) and using \texttt{memcpy} for all (large) collectives.  
    }
    \begin{tabular}{c|rrrr|rrrr}
        \toprule
         GPU & \multicolumn{4}{c|}{FP8} & \multicolumn{4}{c}{BF16} \\  
             &  None & Gather & Scatter & Full & None & Gather & Scatter & Full \\ \midrule
         $4\times \text{4090}$ & 4.3k & 6.3k   &  5.3k   & 7.8k & 2.9k & 4.6k   & 3.6k    & 5.2k\\
         $4\times \text{L40S}$ & 9.5k & 10k    &  9.9k   & 9.9k & 6.8k & 7.0k   & 7.1k    & 6.9k \\ 
         \bottomrule
    \end{tabular}
    \label{tab:nccl-vs-memcpy}
\end{table}

In \autoref{tab:nccl-vs-memcpy}, we show the effect of using \texttt{memcpy}-based communication compared to \texttt{nccl} collectives,
on a system with consumer GPUs without direct peer-to-peer support, and on a professional setup with peer-to-peer capable GPUs.
The data shows that in the consumer setup, \texttt{memcpy} is essential for good performance, whereas there is only a minor difference
on the professional hardware. In that setup, using a combination of \texttt{nccl} and \texttt{memcpy} communication ends up faster
than either option.

\paragraph{DGX Spark Results} Finally, we consider performance on the recent NVIDIA DGX Spark, shown in \autoref{tab:speed-benchmark-spark}. This is interesting because it offers
\SI{128}{\giga \byte} of \emph{unified} memory shared between GPU and CPU, so models up to a size of 7B can fit on a single device without
any offloading. The large memory capacity comes at the cost of low memory bandwidth: at \SI{300}{\giga \byte \per \second},
it is even slower than the 5060Ti's \SI{448}{\giga \byte \per \second}, though still faster than PCIe. However, the lack of faster device memory
altogether means that \emph{any} memory-bound kernel with a working set larger than the L2 cache will operate at this reduced speed, whereas for the consumer
GPUs we are able to mostly hide the PCIe limitations behind explicit prefetching into DRAM.
This also explains why we see little speed-up from switching to FP8 with smaller models: It is precisely these memory-bound kernels that do not
benefit from FP8, and the additional transposes and conversions just add more memory transfers. Only at 7B  do the matrix multiplications make up
enough of the total time for reduced-precision to provide substantial benefits.
Surprisingly, even in pure BF16, the Spark achieves less MFU than the 5060Ti, despite the latter having to rely on activation checkpointing. 
This is probably due to power limits preventing the system from reaching its advertised performance.

\vspace{-1em}
\section{End-to-end results}

\paragraph{Scenario 1: 1.5B pre-training}
Frist, consider pre-training a 1.5B parameter Qwen-style model, to be trained on 4 $\times$ RTX 4090. As training data, we use a retokenized and subsampled version of ClimbMix~\citep{diao2025climb} with 10B tokens and 30B tokens. Due to the long training time involved, we only provide the FP8 version of the 30B run. 
\begin{wrapfigure}{r}{0.5\textwidth}
    \centering
    \includegraphics[width=\linewidth]{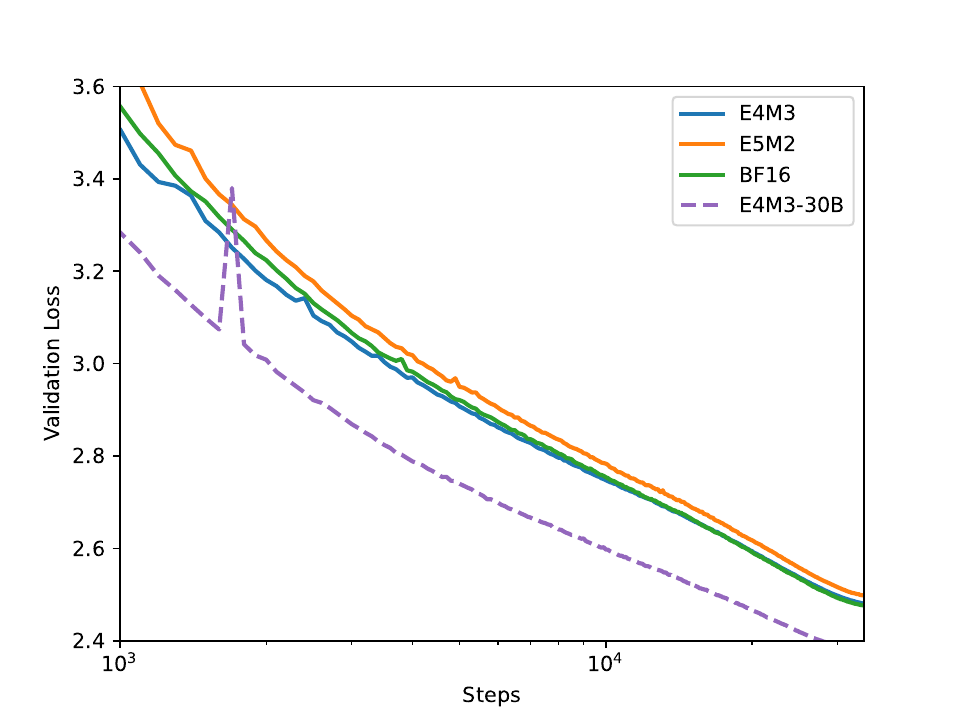}
    \caption{Validation loss vs \# optimizer steps for FP8 and BF16 training on 10B tokens (solid lines), as well as 30B tokens (dashed) for FP8. The 30B run uses a 3x larger batch size.}
    \label{fig:1.5b-training}
\end{wrapfigure}
The results are presented in \autoref{fig:1.5b-training}. We can see that the E4M3 (forward and backward) curve closely tracks that of BF16. In contrast, using E5M2 for activation gradients during backward leads to a minor degradation in performance, in contrast to traditional recommendations~\citep{sun2019hybrid,micikevicius2022FP8,fishman2025scaling}.

\paragraph{Scenario 2: GSM8k}
As a second test, we fine-tune LLama2-7B~\citep{touvron2023llama} and Qwen2.5-14B~\citep{qwen2025qwen25technicalreport} on the GSM8k\citep{cobbe2021gsm8k} dataset. As LLama2 has not been specifically pre-trained with math data, this provides a good demonstration that the performance of LLMs can be improved substantially on narrow domains even with modest computational resources. On the other hand, Qwen models have deliberately used large amounts of mathematical reasoning data during their pre-training, so it is already very good in a few-shot setting, but fails in the zero-shot setting, where performance can be mostly recovered through fine-tuning. Both BF16 and FP8 fine-tuning are able to recover performance.

The results can be seen in \autoref{tab:gsm8k-results}. Training in FP8 does not result in noticable performance degradation compared to the BF16 baseline; On the other
hand, FP8 QAT does confer consistent benefits when inference is also run in FP8. Most notably, while Qwen2.5-14B has already excellent performance in the direct five-shot setting, and receives no benefit from further fine-tuning, there is a significant gap for FP8 inference in the original model that is close with 
 additional fine-tuning.

\begin{table}
    \caption{Performance of LLama2-7B on GSM8K, comparing the original pre-trained model against fine-tuned (T) in BF16 and in FP8, while running inference (I) in either BF16 or FP8. Shown is the average over five independent runs, as well as the standard deviation.}
    \centering
    \begin{tabular}{lr|l|cc|cc|cc}
        \toprule
       Model$\downarrow$ & T$\rightarrow$ & n- & \multicolumn{2}{c|}{Pretrained} &  \multicolumn{2}{c|}{LLMQ BF16} & \multicolumn{2}{c}{LLMQ FP8} \\
        & I$\rightarrow$ & shot & BF16   & FP8     & BF16   & FP8    & BF16    & FP8     \\ \midrule
        \multicolumn{2}{l|}{LLama2-7B}   & 5 & 13.5 & 13.7  & $34.5 \pm 4.3$ & $34.4\pm4.2$ & $37.6 \pm 1.6$  & $37.2 \pm 1.1$  \\
        \multicolumn{2}{l|}{LLama2-7B}   & 0 & \phantom{1}6.0 & \phantom{1}4.9 & $35.7 \pm 1.5$ & $35.4 \pm 1.6$ & $36.6 \pm 0.9$ & $36.4 \pm 2.0$ \\
        \multicolumn{2}{l|}{Qwen2.5-14B} & 5 & 84.1 & 80.7 & $84.5 \pm 0.9$ & $83.5 \pm 0.7$ & $84.0 \pm 0.6$ & $83.9 \pm 0.3$ \\
        \multicolumn{2}{l|}{Qwen2.5-14B} & 0 & 14.0 & 17.8 & $80.4 \pm 1.1$ & $79.1 \pm 0.7$ & $82.7 \pm 0.6$ & $83.1 \pm 0.5$ \\
             \bottomrule
    \end{tabular}
    \label{tab:gsm8k-results}
\end{table}

\section{Limitations}
It is often observed that low-precision works successfully shorter training runs, but breaks down when scaled up~\citep{fishman2025scaling}. However, the intended use case for the software described in this paper are heavily compute-constrained environments, where we do not expect training to run far beyond the Chinchilla-optimal~\citep{hoffmann2022training} ratio; e.g., for the 1.5B model, training 200B tokens in FP8 on 4 RTX 4090 would take about 5 weeks.

\section*{Acknowledgments}

We would like to thank contacts at NVIDIA (Vartika Singh, Nina Carrejo, Kyla Wilkes, and Tijmen Blankevoort), HP (Curtis Burkhalter), and Datacrunch/Verda (Paul Chang and Antonio Dominguez) for hardware support that was essential to this project. 
ES was supported in part by ERC Proof-of-Concept grant FastML. 

\bibliography{reference}

@article{diao2025climb,
  author    = {Shizhe Diao and Yu Yang and Yonggan Fu and Xin Dong and Dan Su and Markus Kliegl and Zijia Chen and Peter Belcak and Yoshi Suhara and Hongxu Yin and Mostofa Patwary and Celine Lin and Jan Kautz and Pavlo Molchanov},
  title={CLIMB: CLustering-based Iterative Data Mixture Bootstrapping for Language Model Pre-training}, 
  journal   = {arXiv preprint},
  year      = {2025},
  archivePrefix = {arXiv},
  primaryClass = {cs.CL},
  url={https://arxiv.org/abs/2504.13161}, 
}

@inproceedings{rajbhandari2020zero,
  title={Zero: Memory optimizations toward training trillion parameter models},
  author={Rajbhandari, Samyam and Rasley, Jeff and Ruwase, Olatunji and He, Yuxiong},
  booktitle={SC20: International Conference for High Performance Computing, Networking, Storage and Analysis},
  pages={1--16},
  year={2020},
  organization={IEEE}
}

@article{hu2022lora,
  title={Lora: Low-rank adaptation of large language models.},
  author={Hu, Edward J and Shen, Yelong and Wallis, Phillip and Allen-Zhu, Zeyuan and Li, Yuanzhi and Wang, Shean and Wang, Lu and Chen, Weizhu and others},
  journal={ICLR},
  volume={1},
  number={2},
  pages={3},
  year={2022}
}

@inproceedings{rajbhandari2021zero,
  title={Zero-infinity: Breaking the gpu memory wall for extreme scale deep learning},
  author={Rajbhandari, Samyam and Ruwase, Olatunji and Rasley, Jeff and Smith, Shaden and He, Yuxiong},
  booktitle={Proceedings of the international conference for high performance computing, networking, storage and analysis},
  pages={1--14},
  year={2021}
}

@inproceedings{dettmers2022bit,
title={8-bit Optimizers via Block-wise Quantization},
author={Tim Dettmers and Mike Lewis and Sam Shleifer and Luke Zettlemoyer},
booktitle={International Conference on Learning Representations},
year={2022},
url={https://openreview.net/forum?id=shpkpVXzo3h}
}

@inproceedings{xi2025coat,
title={{COAT}: Compressing Optimizer states and Activations for Memory-Efficient {FP}8 Training},
author={Haocheng Xi and Han Cai and Ligeng Zhu and Yao Lu and Kurt Keutzer and Jianfei Chen and Song Han},
booktitle={The Thirteenth International Conference on Learning Representations},
year={2025},
url={https://openreview.net/forum?id=XfKSDgqIRj}
}

@article{huang2025stable,
  title={Stable-SPAM: How to Train in 4-Bit More Stably than 16-Bit Adam},
  author={Huang, Tianjin and Hu, Haotian and Zhang, Zhenyu and Jin, Gaojie and Li, Xiang and Shen, Li and Chen, Tianlong and Liu, Lu and Wen, Qingsong and Wang, Zhangyang and others},
  journal={arXiv preprint arXiv:2502.17055},
  year={2025}
}

@article{modoranu2024microadam,
  title={Microadam: Accurate adaptive optimization with low space overhead and provable convergence},
  author={Modoranu, Ionut-Vlad and Safaryan, Mher and Malinovsky, Grigory and Kurti{\'c}, Eldar and Robert, Thomas and Richt{\'a}rik, Peter and Alistarh, Dan},
  journal={Advances in Neural Information Processing Systems},
  volume={37},
  pages={1--43},
  year={2024}
}

@inproceedings{zhao2024galore,
title={GaLore: Memory-Efficient {LLM} Training by Gradient Low-Rank Projection},
author={Jiawei Zhao and Zhenyu Zhang and Beidi Chen and Zhangyang Wang and Anima Anandkumar and Yuandong Tian},
booktitle={Forty-first International Conference on Machine Learning},
year={2024},
url={https://openreview.net/forum?id=hYHsrKDiX7}
}

@inproceedings{shazeer2018adafactor,
  title={Adafactor: Adaptive learning rates with sublinear memory cost},
  author={Shazeer, Noam and Stern, Mitchell},
  booktitle={International Conference on Machine Learning},
  pages={4596--4604},
  year={2018},
  organization={PMLR}
}

@article{fp8lm,
      title={FP8-LM: Training FP8 Large Language Models},
      author={Houwen Peng and Kan Wu and Yixuan Wei and Guoshuai Zhao and Yuxiang Yang and Ze Liu and Yifan Xiong and Ziyue Yang and Bolin Ni and Jingcheng Hu and Ruihang Li and Miaosen Zhang and Chen Li and Jia Ning and Ruizhe Wang and Zheng Zhang and Shuguang Liu and Joe Chau and Han Hu and Peng Cheng},
      year={2023},
      eprint={2310.18313},
      archivePrefix={arXiv},
      primaryClass={cs.LG}
}

@article{micikevicius2022fp8,
  title={Fp8 formats for deep learning},
  author={Micikevicius, Paulius and Stosic, Dusan and Burgess, Neil and Cornea, Marius and Dubey, Pradeep and Grisenthwaite, Richard and Ha, Sangwon and Heinecke, Alexander and Judd, Patrick and Kamalu, John and others},
  journal={arXiv preprint arXiv:2209.05433},
  year={2022}
}

@inproceedings{sun2019hybrid,
 author = {Sun, Xiao and Choi, Jungwook and Chen, Chia-Yu and Wang, Naigang and Venkataramani, Swagath and Srinivasan, Vijayalakshmi (Viji) and Cui, Xiaodong and Zhang, Wei and Gopalakrishnan, Kailash},
 booktitle = {Advances in Neural Information Processing Systems},
 editor = {H. Wallach and H. Larochelle and A. Beygelzimer and F. d\textquotesingle Alch\'{e}-Buc and E. Fox and R. Garnett},
 pages = {},
 publisher = {Curran Associates, Inc.},
 title = {Hybrid 8-bit Floating Point (HFP8) Training and Inference for Deep Neural Networks},
 url = {https://proceedings.neurips.cc/paper_files/paper/2019/file/65fc9fb4897a89789352e211ca2d398f-Paper.pdf},
 volume = {32},
 year = {2019}
}

@inproceedings{fishman2025scaling,
    title={Scaling {FP}8 training to trillion-token {LLM}s},
    author={Maxim Fishman and Brian Chmiel and Ron Banner and Daniel Soudry},
    booktitle={The Thirteenth International Conference on Learning Representations},
    year={2025},
    url={https://openreview.net/forum?id=E1EHO0imOb}
}

@article{renee_2023,
  title={Renee: End-to-end training of extreme classification models},
  author={Jain, Vidit and Prakash, Jatin and Saini, Deepak and Jiao, Jian and Ramjee, Ramachandran and Varma, Manik},
  journal={Proceedings of Machine Learning and Systems},
  year={2023}
}

@inproceedings{hsu2025ligerkernel,
title={Liger-Kernel: Efficient Triton Kernels for {LLM} Training},
author={Pin-Lun Hsu and Yun Dai and Vignesh Kothapalli and Qingquan Song and Shao Tang and Siyu Zhu and Steven Shimizu and Shivam Sahni and Haowen Ning and Yanning Chen and Zhipeng Wang},
booktitle={Championing Open-source DEvelopment in ML Workshop {@} ICML25},
year={2025},
url={https://openreview.net/forum?id=36SjAIT42G}
}

@article{touvron2023llama,
  title={Llama 2: Open foundation and fine-tuned chat models},
  author={Touvron, Hugo and Martin, Louis and Stone, Kevin and Albert, Peter and Almahairi, Amjad and Babaei, Yasmine and Bashlykov, Nikolay and Batra, Soumya and Bhargava, Prajjwal and Bhosale, Shruti and others},
  journal={arXiv preprint arXiv:2307.09288},
  year={2023}
}

@article{cobbe2021gsm8k,
  title={Training Verifiers to Solve Math Word Problems},
  author={Cobbe, Karl and Kosaraju, Vineet and Bavarian, Mohammad and Chen, Mark and Jun, Heewoo and Kaiser, Lukasz and Plappert, Matthias and Tworek, Jerry and Hilton, Jacob and Nakano, Reiichiro and Hesse, Christopher and Schulman, John},
  journal={arXiv preprint arXiv:2110.14168},
  year={2021}
}

@inproceedings{hoffmann2022training,
author = {Hoffmann, Jordan and Borgeaud, Sebastian and Mensch, Arthur and Buchatskaya, Elena and Cai, Trevor and Rutherford, Eliza and de Las Casas, Diego and Hendricks, Lisa Anne and Welbl, Johannes and Clark, Aidan and Hennigan, Tom and Noland, Eric and Millican, Katie and van den Driessche, George and Damoc, Bogdan and Guy, Aurelia and Osindero, Simon and Simonyan, Karen and Elsen, Erich and Vinyals, Oriol and Rae, Jack W. and Sifre, Laurent},
title = {Training compute-optimal large language models},
year = {2022},
isbn = {9781713871088},
publisher = {Curran Associates Inc.},
address = {Red Hook, NY, USA},
booktitle = {Proceedings of the 36th International Conference on Neural Information Processing Systems},
articleno = {2176},
numpages = {15},
location = {New Orleans, LA, USA},
series = {NIPS '22}
}

@inproceedings{NEURIPS2019_093f65e0,
 author = {Huang, Yanping and Cheng, Youlong and Bapna, Ankur and Firat, Orhan and Chen, Dehao and Chen, Mia and Lee, HyoukJoong and Ngiam, Jiquan and Le, Quoc V and Wu, Yonghui and Chen, zhifeng},
 booktitle = {Advances in Neural Information Processing Systems},
 editor = {H. Wallach and H. Larochelle and A. Beygelzimer and F. d\textquotesingle Alch\'{e}-Buc and E. Fox and R. Garnett},
 pages = {},
 publisher = {Curran Associates, Inc.},
 title = {GPipe: Efficient Training of Giant Neural Networks using Pipeline Parallelism},
 url = {https://proceedings.neurips.cc/paper_files/paper/2019/file/093f65e080a295f8076b1c5722a46aa2-Paper.pdf},
 volume = {32},
 year = {2019}
}

@inproceedings{qi2024zero,
title={Zero Bubble (Almost) Pipeline Parallelism},
author={Penghui Qi and Xinyi Wan and Guangxing Huang and Min Lin},
booktitle={The Twelfth International Conference on Learning Representations},
year={2024},
url={https://openreview.net/forum?id=tuzTN0eIO5}
}

@inproceedings{liu2024ringattention,
title={RingAttention with Blockwise Transformers for Near-Infinite Context},
author={Hao Liu and Matei Zaharia and Pieter Abbeel},
booktitle={The Twelfth International Conference on Learning Representations},
year={2024},
url={https://openreview.net/forum?id=WsRHpHH4s0}
}

@misc{ultrascale_playbook,
      title={The Ultra-Scale Playbook: Training LLMs on GPU Clusters},
      author={Nouamane Tazi and Ferdinand Mom and Haojun Zhao and Phuc Nguyen and Mohamed Mekkouri and Leandro Werra and Thomas Wolf},
      year={2025},
}

@misc{blackwell,
      title={NVIDIA Blackwell Architecture Technical Brief},
      year={2024},
      url={https://resources.nvidia.com/en-us-blackwell-architecture}
}

@article{chen2016training,
  title={Training deep nets with sublinear memory cost},
  author={Chen, Tianqi and Xu, Bing and Zhang, Chiyuan and Guestrin, Carlos},
  journal={arXiv preprint arXiv:1604.06174},
  year={2016}
}

@article{dao2022flashattention,
  title={Flashattention: Fast and memory-efficient exact attention with io-awareness},
  author={Dao, Tri and Fu, Dan and Ermon, Stefano and Rudra, Atri and R{\'e}, Christopher},
  journal={Advances in neural information processing systems},
  volume={35},
  pages={16344--16359},
  year={2022}
}

@inproceedings{ren2021zero,
  title={$\{$Zero-offload$\}$: Democratizing $\{$billion-scale$\}$ model training},
  author={Ren, Jie and Rajbhandari, Samyam and Aminabadi, Reza Yazdani and Ruwase, Olatunji and Yang, Shuangyan and Zhang, Minjia and Li, Dong and He, Yuxiong},
  booktitle={2021 USENIX Annual Technical Conference (USENIX ATC 21)},
  pages={551--564},
  year={2021}
}

@article{kingma2014adam,
author = {Kingma, Diederik and Ba, Jimmy},
year = {2014},
month = {12},
pages = {},
title = {Adam: A Method for Stochastic Optimization},
journal = {International Conference on Learning Representations}
}

@inproceedings{micikevicius2018mixed,
title={Mixed Precision Training},
author={Paulius Micikevicius and Sharan Narang and Jonah Alben and Gregory Diamos and Erich Elsen and David Garcia and Boris Ginsburg and Michael Houston and Oleksii Kuchaiev and Ganesh Venkatesh and Hao Wu},
booktitle={International Conference on Learning Representations},
year={2018},
url={https://openreview.net/forum?id=r1gs9JgRZ},
}

@misc{micikevicius2022fp8formatsdeeplearning,
      title={FP8 Formats for Deep Learning}, 
      author={Paulius Micikevicius and Dusan Stosic and Neil Burgess and Marius Cornea and Pradeep Dubey and Richard Grisenthwaite and Sangwon Ha and Alexander Heinecke and Patrick Judd and John Kamalu and Naveen Mellempudi and Stuart Oberman and Mohammad Shoeybi and Michael Siu and Hao Wu},
      year={2022},
      url={https://arxiv.org/abs/2209.05433}, 
}

@article{Kalamkar2019ASO,
  title={A Study of BFLOAT16 for Deep Learning Training},
  author={Dhiraj D. Kalamkar and Dheevatsa Mudigere and Naveen Mellempudi and Dipankar Das and Kunal Banerjee and Sasikanth Avancha and Dharma Teja Vooturi and Nataraj Jammalamadaka and Jianyu Huang and Hector Yuen and Jiyan Yang and Jongsoo Park and Alexander Heinecke and Evangelos Georganas and Sudarshan M. Srinivasan and Abhisek Kundu and Mikhail Smelyanskiy and Bharat Kaul and Pradeep K. Dubey},
  journal={ArXiv},
  year={2019},
}

@article{rouhani2023microscaling,
  title={Microscaling data formats for deep learning},
  author={Rouhani, Bita Darvish and Zhao, Ritchie and More, Ankit and Hall, Mathew and Khodamoradi, Alireza and Deng, Summer and Choudhary, Dhruv and Cornea, Marius and Dellinger, Eric and Denolf, Kristof and others},
  journal={arXiv preprint arXiv:2310.10537},
  year={2023}
}

@book{patterson2016computer,
  title={Computer organization and design ARM edition: the hardware software interface},
  author={Patterson, David A and Hennessy, John L},
  year={2016},
  publisher={Morgan kaufmann}
}

@inproceedings{choquette2020nvidia,
  title={Nvidia a100 gpu: Performance \& innovation for gpu computing},
  author={Choquette, Jack and Gandhi, Wish},
  booktitle={2020 IEEE Hot Chips 32 Symposium (HCS)},
  pages={1--43},
  year={2020},
  organization={IEEE Computer Society}
}

@misc{deepseekai2025deepseekv3technicalreport,
      title={DeepSeek-V3 Technical Report}, 
      author={DeepSeek-AI and Aixin Liu and Bei Feng and Bing Xue and Bingxuan Wang and Bochao Wu and Chengda Lu and Chenggang Zhao and Chengqi Deng and Chenyu Zhang and Chong Ruan and Damai Dai and Daya Guo and Dejian Yang and Deli Chen and Dongjie Ji and Erhang Li and Fangyun Lin and Fucong Dai and Fuli Luo and Guangbo Hao and Guanting Chen and Guowei Li and H. Zhang and Han Bao and Hanwei Xu and Haocheng Wang and Haowei Zhang and Honghui Ding and Huajian Xin and Huazuo Gao and Hui Li and Hui Qu and J. L. Cai and Jian Liang and Jianzhong Guo and Jiaqi Ni and Jiashi Li and Jiawei Wang and Jin Chen and Jingchang Chen and Jingyang Yuan and Junjie Qiu and Junlong Li and Junxiao Song and Kai Dong and Kai Hu and Kaige Gao and Kang Guan and Kexin Huang and Kuai Yu and Lean Wang and Lecong Zhang and Lei Xu and Leyi Xia and Liang Zhao and Litong Wang and Liyue Zhang and Meng Li and Miaojun Wang and Mingchuan Zhang and Minghua Zhang and Minghui Tang and Mingming Li and Ning Tian and Panpan Huang and Peiyi Wang and Peng Zhang and Qiancheng Wang and Qihao Zhu and Qinyu Chen and Qiushi Du and R. J. Chen and R. L. Jin and Ruiqi Ge and Ruisong Zhang and Ruizhe Pan and Runji Wang and Runxin Xu and Ruoyu Zhang and Ruyi Chen and S. S. Li and Shanghao Lu and Shangyan Zhou and Shanhuang Chen and Shaoqing Wu and Shengfeng Ye and Shengfeng Ye and Shirong Ma and Shiyu Wang and Shuang Zhou and Shuiping Yu and Shunfeng Zhou and Shuting Pan and T. Wang and Tao Yun and Tian Pei and Tianyu Sun and W. L. Xiao and Wangding Zeng and Wanjia Zhao and Wei An and Wen Liu and Wenfeng Liang and Wenjun Gao and Wenqin Yu and Wentao Zhang and X. Q. Li and Xiangyue Jin and Xianzu Wang and Xiao Bi and Xiaodong Liu and Xiaohan Wang and Xiaojin Shen and Xiaokang Chen and Xiaokang Zhang and Xiaosha Chen and Xiaotao Nie and Xiaowen Sun and Xiaoxiang Wang and Xin Cheng and Xin Liu and Xin Xie and Xingchao Liu and Xingkai Yu and Xinnan Song and Xinxia Shan and Xinyi Zhou and Xinyu Yang and Xinyuan Li and Xuecheng Su and Xuheng Lin and Y. K. Li and Y. Q. Wang and Y. X. Wei and Y. X. Zhu and Yang Zhang and Yanhong Xu and Yanhong Xu and Yanping Huang and Yao Li and Yao Zhao and Yaofeng Sun and Yaohui Li and Yaohui Wang and Yi Yu and Yi Zheng and Yichao Zhang and Yifan Shi and Yiliang Xiong and Ying He and Ying Tang and Yishi Piao and Yisong Wang and Yixuan Tan and Yiyang Ma and Yiyuan Liu and Yongqiang Guo and Yu Wu and Yuan Ou and Yuchen Zhu and Yuduan Wang and Yue Gong and Yuheng Zou and Yujia He and Yukun Zha and Yunfan Xiong and Yunxian Ma and Yuting Yan and Yuxiang Luo and Yuxiang You and Yuxuan Liu and Yuyang Zhou and Z. F. Wu and Z. Z. Ren and Zehui Ren and Zhangli Sha and Zhe Fu and Zhean Xu and Zhen Huang and Zhen Zhang and Zhenda Xie and Zhengyan Zhang and Zhewen Hao and Zhibin Gou and Zhicheng Ma and Zhigang Yan and Zhihong Shao and Zhipeng Xu and Zhiyu Wu and Zhongyu Zhang and Zhuoshu Li and Zihui Gu and Zijia Zhu and Zijun Liu and Zilin Li and Ziwei Xie and Ziyang Song and Ziyi Gao and Zizheng Pan},
      year={2025},
      eprint={2412.19437},
      archivePrefix={arXiv},
      url={https://arxiv.org/abs/2412.19437}, 
}

@inproceedings{panferov2025quest,
title={Qu{EST}: Stable Training of {LLM}s with 1-Bit Weights and Activations},
author={Andrei Panferov and Jiale Chen and Soroush Tabesh and Mahdi Nikdan and Dan Alistarh},
booktitle={Forty-second International Conference on Machine Learning},
year={2025},
url={https://openreview.net/forum?id=I0Ux2nAN6u}
}

@misc{qwen2025qwen25technicalreport,
      title={Qwen2.5 Technical Report}, 
      author={Qwen and : and An Yang and Baosong Yang and Beichen Zhang and Binyuan Hui and Bo Zheng and Bowen Yu and Chengyuan Li and Dayiheng Liu and Fei Huang and Haoran Wei and Huan Lin and Jian Yang and Jianhong Tu and Jianwei Zhang and Jianxin Yang and Jiaxi Yang and Jingren Zhou and Junyang Lin and Kai Dang and Keming Lu and Keqin Bao and Kexin Yang and Le Yu and Mei Li and Mingfeng Xue and Pei Zhang and Qin Zhu and Rui Men and Runji Lin and Tianhao Li and Tianyi Tang and Tingyu Xia and Xingzhang Ren and Xuancheng Ren and Yang Fan and Yang Su and Yichang Zhang and Yu Wan and Yuqiong Liu and Zeyu Cui and Zhenru Zhang and Zihan Qiu},
      year={2025},
      eprint={2412.15115},
      archivePrefix={arXiv},
      primaryClass={cs.CL},
      url={https://arxiv.org/abs/2412.15115}, 
}

@inproceedings{zheng2024llamafactory,
  title={LlamaFactory: Unified Efficient Fine-Tuning of 100+ Language Models},
  author={Yaowei Zheng and Richong Zhang and Junhao Zhang and Yanhan Ye and Zheyan Luo and Zhangchi Feng and Yongqiang Ma},
  booktitle={Proceedings of the 62nd Annual Meeting of the Association for Computational Linguistics (Volume 3: System Demonstrations)},
  address={Bangkok, Thailand},
  publisher={Association for Computational Linguistics},
  year={2024},
  url={http://arxiv.org/abs/2403.13372}
}

@article{korthikanti2023reducing,
  title={Reducing activation recomputation in large transformer models},
  author={Korthikanti, Vijay Anand and Casper, Jared and Lym, Sangkug and McAfee, Lawrence and Andersch, Michael and Shoeybi, Mohammad and Catanzaro, Bryan},
  journal={Proceedings of Machine Learning and Systems},
  volume={5},
  pages={341--353},
  year={2023}
}

@inproceedings{hernandez-cano2025towards,
title={Towards Fully {FP}8 {GEMM} {LLM} Training at Scale},
author={Alejandro Hern{\'a}ndez-Cano and Dhia Garbaya and Imanol Schlag and Martin Jaggi},
booktitle={The Thirty-ninth Annual Conference on Neural Information Processing Systems},
year={2025},
url={https://openreview.net/forum?id=KYTFXxTJ12}
}


\appendix
\section{Appendix}

\subsection{Optimal Configurations}
\autoref{tab:run-config-single-gpu} shows the configuration used to achieve the throughput reported in \autoref{tab:speed-benchmark-single}.
The configurations selected for the LLama-factory baseline runs 
are given in \autoref{tab:run-config-lf}.

\begin{table}
    \caption{Offloading: $x$ - residual, $m, v$ - optimizer moments, $\theta^*, \theta$ - (master) parameters, $g$ - gradient}
    \centering
    \begin{tabular}{rrrrrr}
        \toprule
        GPU & Size & DType & Batch & Recompute & Offload \\ \midrule
        5060Ti & 0.5B &  FP8 & 12 & --- & --- \\
               & 0.5B & BF16 & 10 & --- & --- \\
               & 1.5B &  FP8 & 8  & Block & $x$ \\
               & 1.5B & BF16 & 8  & FFN, Att & --- \\
               & 3B   &  FP8 & 12 & Block & $m$, $v$, $\theta^*$ \\
               & 3B   & BF16 & 12 & QKV, FFN & $m$, $v$, $\theta$ \\
               & 7B   &  FP8 & 32 & Block & $x$, $m$, $v$, $g$, $\theta$, $\theta^*$ \\
               & 7B   & BF16 & 32 & Block & $x$, $m$, $v$, $g$, $\theta$ \\\midrule
        4090   & 0.5B &  FP8 & 16 & ---    & --- \\
               & 0.5B & BF16 & 16 & ---    & --- \\
               & 1.5B & FP8  & 4  & ---    & --- \\
               & 1.5B & BF16 & 4  & ---    & --- \\
               & 3B   & FP8  & 4  & ---    & $m$, $v$, $\theta^*$ \\
               & 3B   & BF16 & 4  & SwiGLU & $m$, $v$ \\
               & 7B   & FP8  & 16 & Block  &  $m$, $v$, $\theta^*$, $\theta$, $x$ \\
               & 7B   & BF16 & 16 & Block &  $m$, $v$, $\theta$, $x$ \\
               & 14B  & FP8  & 32 & Block &  $x$, $m$, $v$, $g$, $\theta$, $\theta^*$ \\
               & 14B & BF16  & 32 & Block &  $x$, $m$, $v$, $g$, $\theta$
        \\ \bottomrule
    \end{tabular}
    \label{tab:run-config-single-gpu}
\end{table}

\begin{table}
    \centering
    \caption{Configurations for LLama-factory. Activation checkpointing is enabled in all settings.}
    \begin{tabular}{r|rr|rr}
        \toprule
        & \multicolumn{2}{c|}{1x4090} &  \multicolumn{2}{c}{4x4090} \\
        Size & Batch & Offload & Batch & Offload \\ \midrule
        0.5B & 128   & ---     & 128 & --- \\
        1.5B & 16    & ---     & 32  & --- \\
        3B   & 48    & ZeRO-2  & 64  & ZeRO-3 \\
        7B   & 32    & ZeRO-3  & 32  & ZeRO-3 \\
        14B  & 20    & ZeRO-3  & 21  & ZeRO-3 \\
        32B  & OOM   & OOM     & 4   & ZeRO-3 \\
        \bottomrule
    \end{tabular}
    \label{tab:run-config-lf}
\end{table}

\subsection{GSM8k hyperparameters}
For LLama2-7B, we train for two epochs at a batch size of \num{32768} tokens per optimizer step with a sequence length of \num{2048} and learning rate \num{2e-5} linearly decaying to 25\% after an initial warm-up of 10 steps.
This training can be completed in about 30~minutes on a single 4090, or 90~minutes on a 5060Ti.

For Qwen2.5-14B, we train for one epoch with a batch size of 96 at sequence length 512 for a single epoch, decaying the learning rate from \num{6e-6} to 0 after 20 warm-up steps.

\subsection{A note on MFU}
\label{app:peak-flops}
At first glance, the MFU values for L40S seem disappointingly low. The \textquote{bad} performance can be explained with the fact that the MFU was calculated using official spec-sheet peak FLOP/s, which appear to be unattainable in the hardware configuration in question. For example, running a large matrix multiplication ($16384 \times 16384 \times 16384$) on a single GPU only achieves about \SI{270}{\tera \flop\per\second}, just $\nicefrac{3}{4}$
of the advertised \SI{362}{\tera \flop\per\second} peak performance.

For the 5060Ti, we found that a single large matmul managed to achieve up to 108\% of the supposed peak FLOP/s, so the MFUs are slight overestimates. In case of the 4090, we similarly benchmark around $103\%$ of peak performance.

For the DGX Spark, similar to the L40S, we see about 70\% of peak flops in a matmul micro-benchmark.

\end{document}